\documentclass{ieeeaccess}
\usepackage{cite}
\usepackage{amsmath,amssymb,amsfonts}
\usepackage{algorithmic}
\usepackage[linesnumbered,ruled,vlined]{algorithm2e}
\usepackage[pdftex]{graphicx}
\usepackage{textcomp}
\usepackage{array}
\usepackage{url}
\usepackage{dirtytalk}
\usepackage[table,xcdraw]{xcolor}
\usepackage[linesnumbered,ruled,vlined]{algorithm2e}
\usepackage{bm}
\usepackage{multirow}
\usepackage{mathtools}
\usepackage{float}
\usepackage{tabularx}
\usepackage{amssymb}

\usepackage{booktabs}
\usepackage{multirow}
\usepackage[normalem]{ulem}
\useunder{\uline}{\ul}{}

\usepackage[ruled,linesnumbered]{algorithm2e}

\SetKwInput{KwInput}{Input}                
\SetKwInput{KwOutput}{Output}              

\def\BibTeX{{\rm B\kern-.05em{\sc i\kern-.025em b}\kern-.08em
    T\kern-.1667em\lower.7ex\hbox{E}\kern-.125emX}}
\begin{document}
\history{}
\doi{}

\title{Distributed Video Adaptive Block Compressive Sensing }
\author{\uppercase{Joseph Zammit}\authorrefmark{1}, \IEEEmembership{Member, IEEE},
\uppercase{Ian J. Wassell\authorrefmark{1}}.
\IEEEmembership{Member, IET}}
\address[1]{Computer Laboratory, William Gates Building, 15 JJ Thomson Ave, Cambridge CB3 0FD, United Kingdom}
\tfootnote{The authors would like to thank and acknowledge funding from  the  Cambridge  Trust,  the  Engineering  and  Physical Sciences  Research  Council  Centre  for  Doctoral  Training  in Sensor  Technologies  and  Applications  (EP/L015889/1), and the Endeavour (Malta) Scholarships Scheme.}


\corresp{Corresponding author: Joseph Zammit (e-mail: jz390@cl.cam.ac.uk).}

\begin{abstract}
Video block compressive sensing has been studied for use in resource constrained scenarios, such as wireless sensor networks, but the approach still suffers from low performance and long reconstruction time. Inspired by classical distributed video coding, we design a lightweight encoder with computationally intensive operations, such as video frame interpolation, performed at the decoder. Straying from recent trends in training end-to-end neural networks, we propose two algorithms that leverage convolutional neural network components to reconstruct video with greatly reduced reconstruction time. At the encoder, we leverage temporal correlation between frames and deploy adaptive techniques based on compressive measurements from previous frames. At the decoder, we exploit temporal correlation by using video frame interpolation and temporal differential pulse code modulation. Simulations show that our two proposed algorithms, VAL-VFI and VAL-IDA-VFI reconstruct higher quality video, achieving state-of-the-art performance.
\end{abstract}

\begin{keywords}
Distributed Video Compressive Sensing, Adaptive Block Compressive Sensing
\end{keywords}

\titlepgskip=-15pt

\maketitle
\section{Introduction}
\IEEEPARstart{T}{ransmitting} video continuously from resource constrained sensors is very challenging because of the need to capture, compress and transmit video as power efficiently as possible, video being one of the highest bit rate signals possible. Classical distributed video coding (DVC) techniques assume implicitly that capturing video is power-efficient, and focus on compressing video as efficiently as possible, so that it can be transmitted viably from the sensor. This means that computationally intensive operations, such as motion compensated prediction (MCP), have to be moved to the decoder. This would lead to a rise in transmission bitrate, and is countered by transmitting some of the frames at a lower bit rate, normally resulting in lower visual quality at the decoder. However, we can leverage the correlation between frames to restore video quality. The distributed source coding theories of Slepian and Wolf \cite{1973-slepian} and Wyner and Ziv \cite{1976-wyner}, inform us that we can transmit correlated data without exploiting correlation at the encoder, at a lower rate than that required if we assume the data is uncorrelated, and still recover the full information at the decoder.

\Figure[h!](topskip=0pt, botskip=0pt, midskip=0pt)[width=0.475\textwidth]{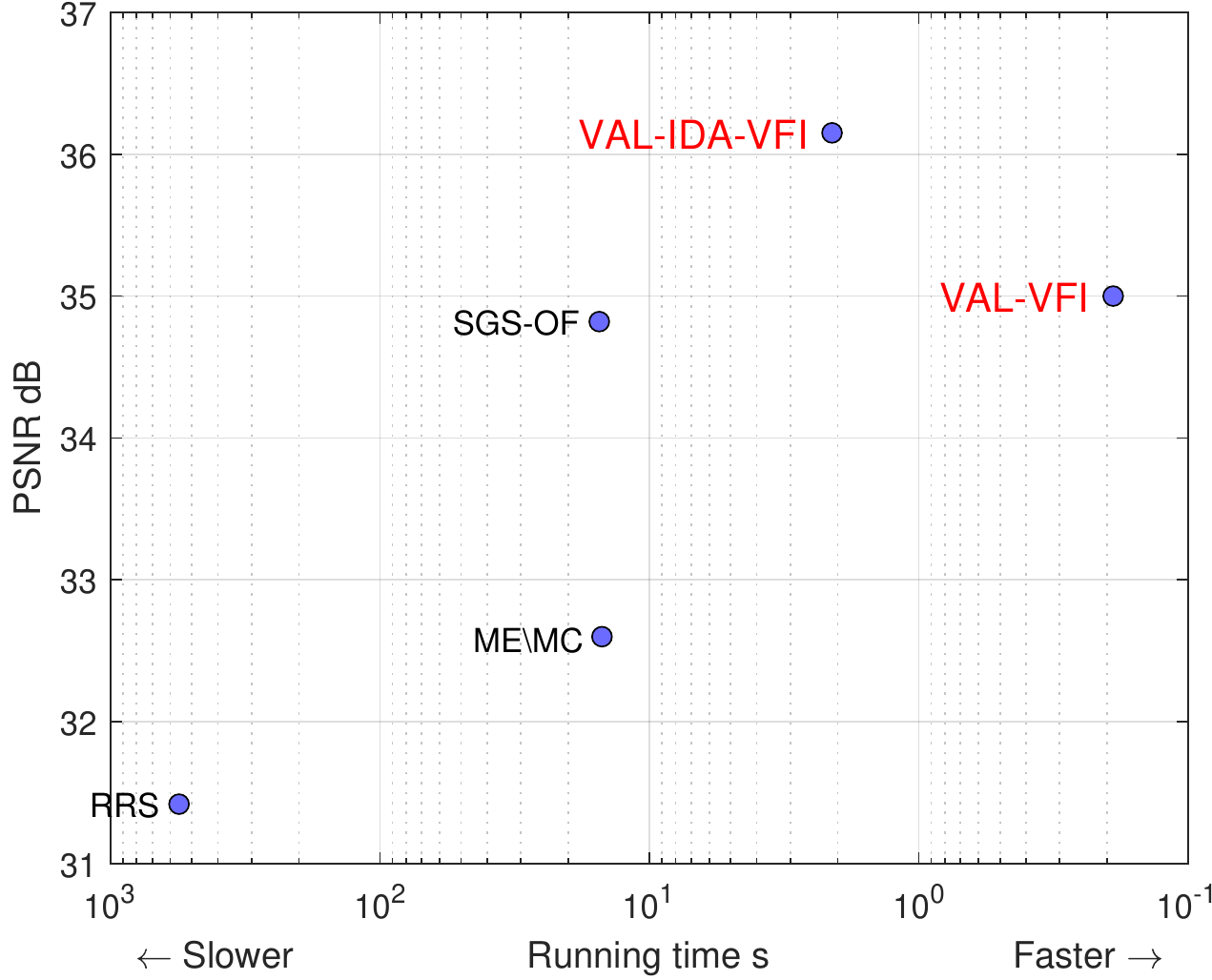}
    {PSNR (dB) versus execution time per frame (s) for Video CS algorithms, calculated for the first 17 frames of six CIF video sequences described in section \ref{VAL-Results}. SGS-OF \cite{2020-chen} PSNR and execution time are reported from the original paper, that employed a comparable simulation platform. \label{PSRN-vs-Time}}
    
Compressive sensing (CS) \cite{2006-Candes}, \cite{2006-Donoho} represents a departure from the normal source coding paradigm, of sampling an analogue signal at the Nyquist rate, digitizing the samples at the highest possible signal-to-quantization-noise-ratio and then using complex source coding algorithms to compress the data as much as possible. In CS, the capture and compress processes are combined to obviate the need for computationally-intensive source coding algorithms. However, using information theoretic arguments, Goyal \cite{2008-goyal} cautions that the compression factors achievable with CS alone, are lower than those that can be achieved by classical source coding paradigms. This means that some form of low-complexity source coding is still required prior to transmission.

CS senses a multidimensional signal $\mathbf{x} \in \mathbb{R}^N$ by performing the inner product between all the components of vector $\mathbf{x}$ and rows of a measurement matrix $\mathbf{\Phi} \in \mathbb{R}^{M \times N}$. The CS operation generates an $M$ dimensional compressed signal $\mathbf{y = \Phi x}$, that is $\mathbf{y} \in \mathbb{R}^M$. This reduces the dimensionality of $\mathbf{x}$ from N to the dimensionality of $\mathbf{y}$, achieving a compression ratio, $\delta = M/N$. Note that a higher compression ratio means more measurements.

Applying CS to video frames requires the storage of the measurement matrix $\mathbf{\Phi} \in \mathbb{R}^{M \times N}$ at the encoder. Contemporary video CS techniques operate on a video signal with a group of pictures (GOP) structure. The leading key frame in the GOP is coded with a higher quality than the remaining non-key frames. The key frame is typically coded with a compression ratio $\delta$ in the range 0.4 to 0.7 in the literature \cite{2010-Mun} \cite{2016-zhao} \cite{2020-chen}. This means that even for modest resolution, common intermediate format (CIF) video, that is with $N=352\times288$ pixels per frame, the storage requirement of the measurement matrix $N_s=\delta \cdot N^2$ is of the order of gigabytes when representing each matrix coefficient by an integer (e.g., 16 bits) and is impractical for resource constrained sensors. Generating the matrix on the fly would be too power inefficient. The solution to this problem, proposed in the literature \cite{2007-Gan}, \cite{2012-Fowler}, is to break the video frames into sub-image blocks of size $B \times B$ pixels, where B is typically 4, 8, 16 or 32. The same measurement matrix is then used to compressively sense each block. 

However, block-based CS of video frames comes with two problems. The first arises from the fact that the number of measurements $M$ required by CS theory is proportional to the sparsity of the signal $S$, that is $M=kS\log(N/S)$, where $k$ is a constant with typical values in the range \{$2\dots6$\} \cite{baraniuk2008simple}. The sparsity is the number of non-zero coefficients, in some domain $\mathbf{\Psi}$, that can represent the signal with the desired quality. If the video frame is processed as one block, there is one sparsity value for the frame. With block-based processing, each block has its own sparsity level, that is changing from frame to frame. In adaptive block-based image CS, the sparsity of each block is estimated prior to compressively sensing it \cite{2020-zammit}. The challenge is to estimate the sparsity with as low a complexity and overhead as possible.

The second problem caused by block-based coding, is that of blocking artefacts appearing in the reconstructed image, if the CS reconstruction is also block-by-block. This can be solved by applying deblocking filters or by sensing the image block-by-block but reconstructing it as a whole, using a full-image sensing matrix that is composed of the block sensing matrices on its diagonal, that is $\mathbf{\Phi_I} = diag(\mathbf{\Phi_B})$ \cite{2007-Gan}.

In the following, bold capital letters represent matrices, bold lower case letters represent vectors, and normal case letters, scalar values. Consider a multidimensional signal $\mathbf{x} \in \mathbb{R}^N$ with a sparse representation $\mathbf{f} \in \mathbb{R}^N$ in some domain $\mathbf{\Psi}$, that is $\mathbf{f = \Psi x}$. Then the compressive samples of $\mathbf{y}$, produced by the measurement matrix $\mathbf{\Phi}$ are given by $\mathbf{y = \Phi\Psi x + n}$ where $\mathbf{n}$ is the measurement noise. The reconstruction of $\mathbf{x}$ requires the solution of this equation. However, this is an ill-posed problem because the sensing matrix $\mathbf{A := \Phi\Psi} \in \mathbb{R}^{M \times N}$ and $M<<N$. A tractable solution can be pursued by first casting the problem as a convex linear program:
\begin{equation} \label{BPDN}
    \min_{\mathbf{x}} ~~~ {\lVert \mathbf{x} \rVert}_1, ~~~ \mathbf{s.t. ~~{\lVert Ax - y \rVert_2^2 < \epsilon} }
\end{equation}
where $\epsilon$ is a measure of the noise level, and solving it using state-of-the-art LP solvers. This is nontrivial and time consuming, and a significant number of reconstruction techniques have been proposed to accelerate the solution, such as matching pursuit \cite{1993-mallat}, Bayesian \cite{2009-baron}, approximate message passing (AMP) \cite{2009-Donoho} and denoising AMP (D-AMP) \cite{2016-Metzler}. 

Recently, convolutional neural networks (CNNs) have been applied to solve equation (\ref{BPDN})  \cite{2015-mousavi}, \cite{2017-metzler}, \cite{2018-Zhao}, \cite{2020-zammit}. A number of authors have proposed CNNs that can reconstruct compressively sensed images in 10's to 100's of milliseconds, such as Reconnet \cite{2016-kulkarni}, CS-Net \cite{2017-shi}, ISTA-Net and ISTA-Net+ \cite{2018-zhang}, and SCSNet \cite{2019-shi}. The reconstruction times of these CNNs allows CS video to be transmitted at tens of frames per second, but without exploiting temporal correlation between frames, generating large bit-rates.

Video CS techniques have been proposed that exploit temporal correlation by multihypothesis prediction, motion compensated prediction or optical flow at the decoder such as ME/MC \cite{2010-Mun}, RRS \cite{2016-zhao}, and SGS-OF \cite{2020-chen}. Recently video CS (VCS) has been reconstructed using CNNs, for example VCSNet\cite{2020a-shi}.

In this paper we present a real-time video compressive sensing framework that leverages plug-and-play CNNs to exploit temporal correlation between frames at the decoder, using video-frame interpolation (VFI). The DAIN \cite{2019-bao} VFI CNN allows our VCS adaptive linear DCT VFI (VAL-VFI) algorithm, shown in figure \ref{PSRN-vs-Time}, to achieve state-of-the-art PSNR and MS-SSIM \cite{2003-wang} performance. The main contributions of this paper are as follows:

\begin{figure*}[t!]
    \centering
    \includegraphics[width=0.9\textwidth]{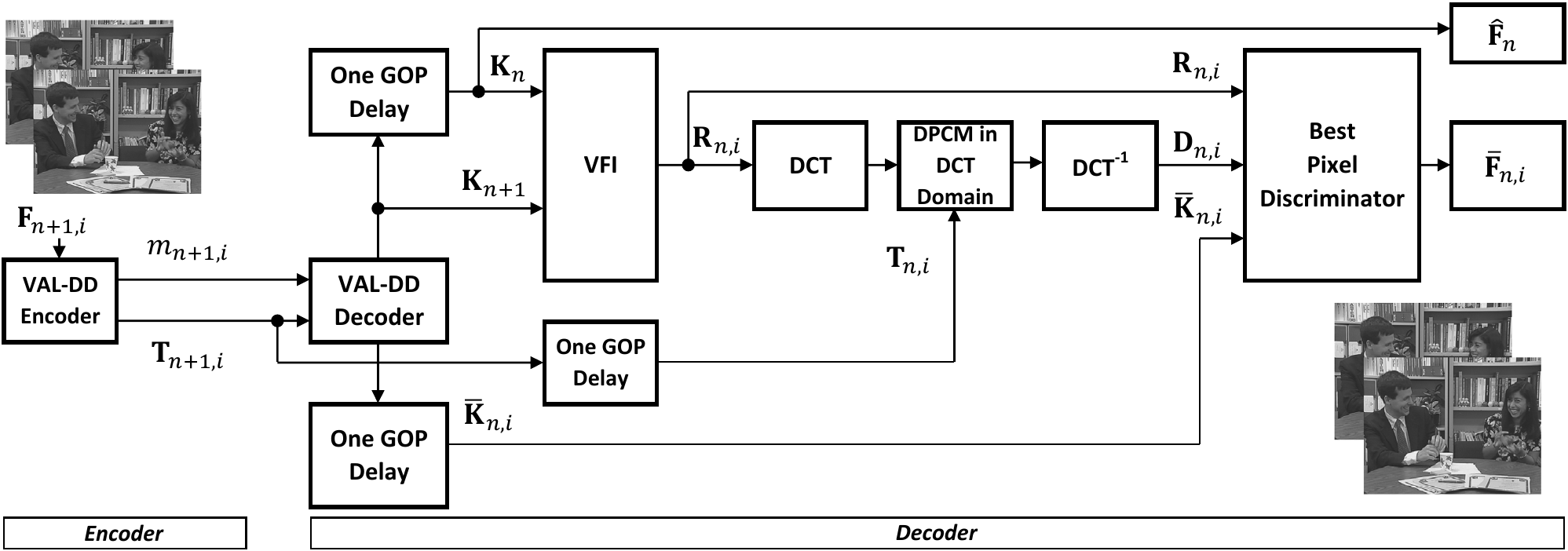}
    \caption{Block diagram of the distributed VAL-VFI algorithm. The VAL-DD decoder has optional IDA reconstruction.}
    \label{VAL-VFI-fig}
\end{figure*}

\begin{itemize}
  \item   In a departure from the current focus to design decoders that execute solely on the GPU, we design hybrid decoders that use the GPU to accelerate two computationally intensive components of our proposed algorithms, video frame interpolation and full image reconstruction.
  \item  We propose a video CS framework using adaptive linear DCT measurements (VAL) that exploits temporal correlation at the encoder and the decoder. At the encoder, two algorithms, THI and MDD are proposed to leverage temporal correlation by adapting the block measurements based on the transform coefficients measured in previous frames. 
  \item We prove that using a linear 2D-DCT measurement matrix allows temporal DPCM at the decoder by simply filtering and mixing transform coefficients from the key and non-key frames.
  \item Coupled with the above, and using a plug-and-play, CNN-based video frame interpolation module, we reach state-of-the-art PSNR and MS-SSIM performance, with an execution time that is two orders of magnitude lower that current state-of-the-art methods. 
  \item Additionally, we can reconstruct the video using the recently proposed iterative denoising algorithm (IDA) \cite{2020-zammit} that can compressively reconstruct a full image from the adaptively and deterministically sensed blocks, to improve the video quality.
\end{itemize}

\section{Related work} \label{Related-Work}
Fowler and Mun \cite{2011-mun} proposed the MC-BCS-SPL algorithm which incorporates block-based motion estimation (ME) and compensation (MC) at the decoder. In a GOP with $P$ pictures, they recover the first $P/2$ frames with forward reconstruction starting from the intra-frame, and the remaining frames, by reconstructing them from the reference frame in the next GOP. They then refine the frames by bi-directional prediction from the previous and following frames. Chen et al. \cite{2011-Chen} took inspiration from the this framework and proposed multi-hypothesis predictions (BCS-SPL-MH) as an improvement for both still images and video. An initial reconstruction is performed using standard BCS-SPL \cite{2011-mun}, and weighted Tikhonov regularization is then used to form predictions of the residual to iteratively improve the reconstruction.

Recently, Zhao et al. \cite{2018-Zhao} proposed a two-phase hybrid intra-frame, inter-frame algorithm. The first phase exploits spatial correlation to produce high-quality reference frames. In the second, the reweighted sparsity of the residual difference between frames is recovered using an algorithm based on split Bregman iteration. Experimental results are presented for the first 16 frames of six popular common CIF video sequences. The GOP consists of an intra-frame followed by seven inter-frames. Block-based sensing is assumed using Gaussian random projections on $32\times32$ pixel blocks. Key frames are sensed at a compression ratio of 0.7 and non-key frames at 0.2. Multi-hypothesis prediction is used to reconstruct the inter-frames. The proposed reweighted residual sparsity (RRS) scheme is compared against four representative video CS reconstruction methods, and was shown to achieve state-of-the-art performance. However, the reconstruction time per frame was reported to exceed 5 minutes.

SGS-OF \cite{2020-chen} is closest to our work, both in terms of concept and results. The authors adopt a GOP with size 8 and compressively sense key and non-key frames with compression ratios 0.7 and 0.1. They propose a structural group sparsity (SGS) model and employ the alternating direction method of multipliers (ADMM) variant of the augmented Lagrangian method (ALM), to reconstruct the independent frames, and then use the Coarse-to-Fine optical flow (OF) method to create a fused image using the forward and backward OF motion estimates. Finally they use the BCS-SPL \cite{2009-mun} algorithm  to refine the original reconstruction using residual estimation and compensation. The authors investigate both Gaussian and partial DCT measurement matrices and report large gains with the deterministic Partial DCT matrices. The authors quote a reconstruction time of around 15s per frame using the partial DCT sensing matrix.


\section{Distributed video compressive sensing}\label{VAL-VFI}
\begin{figure}[t!]
    \centering
    \includegraphics[width=0.45\textwidth]{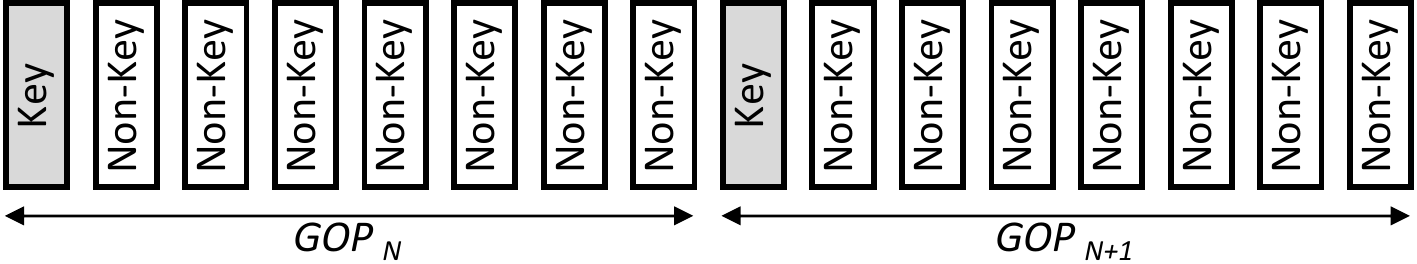}
    \caption{GOP Structure. N is the Group of Pictures number.}
    \label{GOP-MID}
\end{figure}

Inspired by distributed video compressive sensing techniques, we develop the VAL framework described in this section and evaluate it empirically in section \ref{VAL-Results}. The block diagram of the proposed VAL-VFI algorithm is shown in figure \ref{VAL-VFI-fig}. The GOP structure in figure \ref{GOP-MID} is adopted wherein key frames are sensed with a high average key compression ratio $\delta_K = M/N$ so that the reconstructed quality is high, where $N$ is the number of pixels and $M$ the number of measurements. Non-key frames are sensed with a substantially lower $\delta_{\bar{K}}$. We place the key-frames at the start of the GOPs and then predict the non-key-frames from the key-frames in the current and following GOP, using VFI.

The encoder captures frames $\mathbf{F}_{n+1,i}$ using two adaptive VAL-DD algorithms inspired by the AL-DCT-DD algorithm in \cite{2020-zammit}. $\mathbf{F}_{n+1}$ is the Key frame in GOP $n+1$. $\mathbf{F}_{n+1,i}$ is the $i$-th non-key frame in GOP $n+1$, $i=2..g$ where $g$ is the GOP size. The encoder then transmits the compressively sensed 2D-DCT transform coefficients $\mathbf{T}_{n+1,i}$ and the number of transform coefficients per block $m_{n+1,i}$ sensed in the $i$-th frame of the GOP. 

The decoder decodes the key and non-key frames, in real-time, using the inverse 2D DCT in the VAL-DD decoder, and buffers one GOP worth of transform coefficients and reconstructed non-key frames. This is necessary because the VFI algorithm requires the key frame in the current GOP $\mathbf{K}_n$ and that in the following GOP $\mathbf{K}_{n+1}$, besides the $i$-th non-key frame $\mathbf{\bar{K}}_{n,i}$ in the current GOP.

The VFI block uses the two key frames $\mathbf{K}_n$ and $\mathbf{K}_{n+1}$ to interpolate the non-key frames $\mathbf{R}_{n,i}$ in between. The decoder then computes the temporal DPCM in the transform domain to predict the received frame from the reference frame $\mathbf{R}_{n,i}$ and non-key transform coefficients $\mathbf{T}_{n,i}$. The DPCM output $\mathbf{D}_{n,i}$ is then compared with $\mathbf{R}_{n,i}$ in the \textit{best pixel discriminator} block that outputs $\mathbf{D}_{n,i}$ or $\mathbf{\bar{K}}_{n,i}$ to produce the reconstructed non-key frame $\mathbf{\bar{F}}_{n,i}$. The reconstructed key frame $\mathbf{\hat{F}}_n$ is equals to $\mathbf{K}_n$.

Better quality can be achieved if the VAL-DD reconstruction is carried out using the IDA algorithm with the DnCNN denoiser \cite{2020-zammit}.

\begin{algorithm} [!t]
\small

\KwInput{Image $\mathcal{I}$, compression factor $C_F = N/M$, block size $B$, image width $W$, image height $H$.}

\KwOutput{Number of measurements $m_i$ per block $i$, $m_i$ measurements per block.}

\BlankLine

Crop and partition, $\mathcal{I}$ into $n_B = \lfloor H/B \rfloor \cdot \lfloor W/B \rfloor$, $B \times B$ blocks;

Calculate the number of non-adaptive, phase-1 measurements per block $m_i^1 = \lfloor B^2/(2 \cdot C_F) \rfloor$;

Collect $m_i^1$ low-pass 2D DCT coefficients in zigzag order from each block;

Sort all phase-1 coefficients in descending order of absolute magnitude;

Set a threshold $T_I$ which is equal to the absolute value of the $(M/4)$-th largest coefficient in the sorted list of coefficients, where $M = (H \times W) / C_F$;

Calculate $m_i^T$, the number of coefficients in each block whose absolute value exceeds $T_I$;

Set the number of coefficients to collect from each block in phase 2, $m_i^2$, to  $2 \times m_i^T$;

Collect $m_i^2$ additional coefficients as required;

Output $m_i = m_i^1 + m_i^2$, the number of measurements per block;

Output the $m_i$ measurements per block.

\caption{THI}
\label{Algorithm-THI}
\end{algorithm}

\begin{algorithm} [!t]
\small
 
\KwInput{Image $\mathcal{I}$, transform coefficients of reference image $TC_{Ref}$, compression factor $C_F = N/M$, block size $B$, image width $W$, image height $H$.}

\KwOutput{Number of measurements $m_i$ per block $i$, $m_i$ measurements per block.}

\BlankLine

Crop and partition $\mathcal{I}$ into $n_B = \lfloor H/B \rfloor \cdot \lfloor W/B \rfloor$, $B \times B$ blocks;

Calculate the number of non-adaptive measurements per block $m_i^1 = \lfloor B^2/(2 \cdot C_F) \rfloor$;

Collect $m_i^1$ low-pass 2D DCT coefficients in zigzag order from each block;

Replace the $m_i^1$ low pass coefficients in $TC_{Ref}$ by the TCs collected in step 3;

Set $m_i^2$ equal to the number of largest $M$ transform coefficients in $TC_{Ref}$ by absolute value, that fall in each block $i$;

Collect $m_i^2$ additional coefficients as required;

Output $m_i = m_i^1 + m_i^2$, the number of measurements per block;

Output the $m_i$ measurements per block.

\caption{MDD}
\label{Algorithm-MDD}
\end{algorithm}

\subsection{Adaptive block compressive sensing}

At the encoder, transform coefficients (TC) are sensed in two phases. Measurements from the first phase are used to estimate the number of adaptive measurements in the second phase. We propose two improved versions of the AL-DCT-DD algorithm in \cite{2020-zammit}; threshold over the whole image (THI) and mixed-mode DCT domain (MDD). 

THI is defined in algorithm \ref{Algorithm-THI}. It collects half the TCs equally from all blocks in phase one. These are then used to estimate the number of phase-2 coefficients based on the proportion of the largest phase-1 coefficients from all blocks in the image that fall in the current block.

When compressively sensing successive non-key frames in a video sequence, one would have the benefit of having collected significantly more TCs per block from the key frame, which is sensed at a substantially higher rate. We thus propose another algorithm that adapts the phase-two measurements based on the reference key-frame phase-1 TCs, but replacing the low-pass TCs with the phase-1 TCs of the current frame. We refer to this as the mixed-mode DCT domain (MDD) as described in algorithm \ref{Algorithm-MDD}).

\subsection{IDA reconstruction}
The IDA algorithm proposed in \cite{2020-zammit} can reconstruct adaptive block based images sensed using the deterministic 2D DCT measurement matrix. We propose to use IDA to increase the PSNR and MS-SSIM of the reconstructed video. The IDA algorithm is an iterative thresholding algorithm that uses the CNN-based, DnCNN denoiser \cite{2017-zhang}.

\subsection{Video frame interpolation}
VFI has been extensively studied over the past decades to up-sample the frame rate of video content to match the ever increasing video monitor frame rates and produce smoother playback and slow-motion effects. Recently, CNNs have been leveraged to generate interpolated frames in real-time \cite{2019-nah}. One of the better performing CNN-based systems is DAIN \cite{2019-bao} which can interpolate frames in real-time. Key frames are encoded with a higher quality (i.e., with lower compression) than non-key frames. Therefore, we leverage video frames interpolated from two key frames by DAIN \cite{2019-bao} as high-quality estimates of non-key frames. We then use the VFI output frame $\mathbf{R}_{n,i}$ as the input to the temporal DPCM block, which continues to improve video quality.

\begin{figure}[!t]
    \centering
    \includegraphics[width=0.48\textwidth]{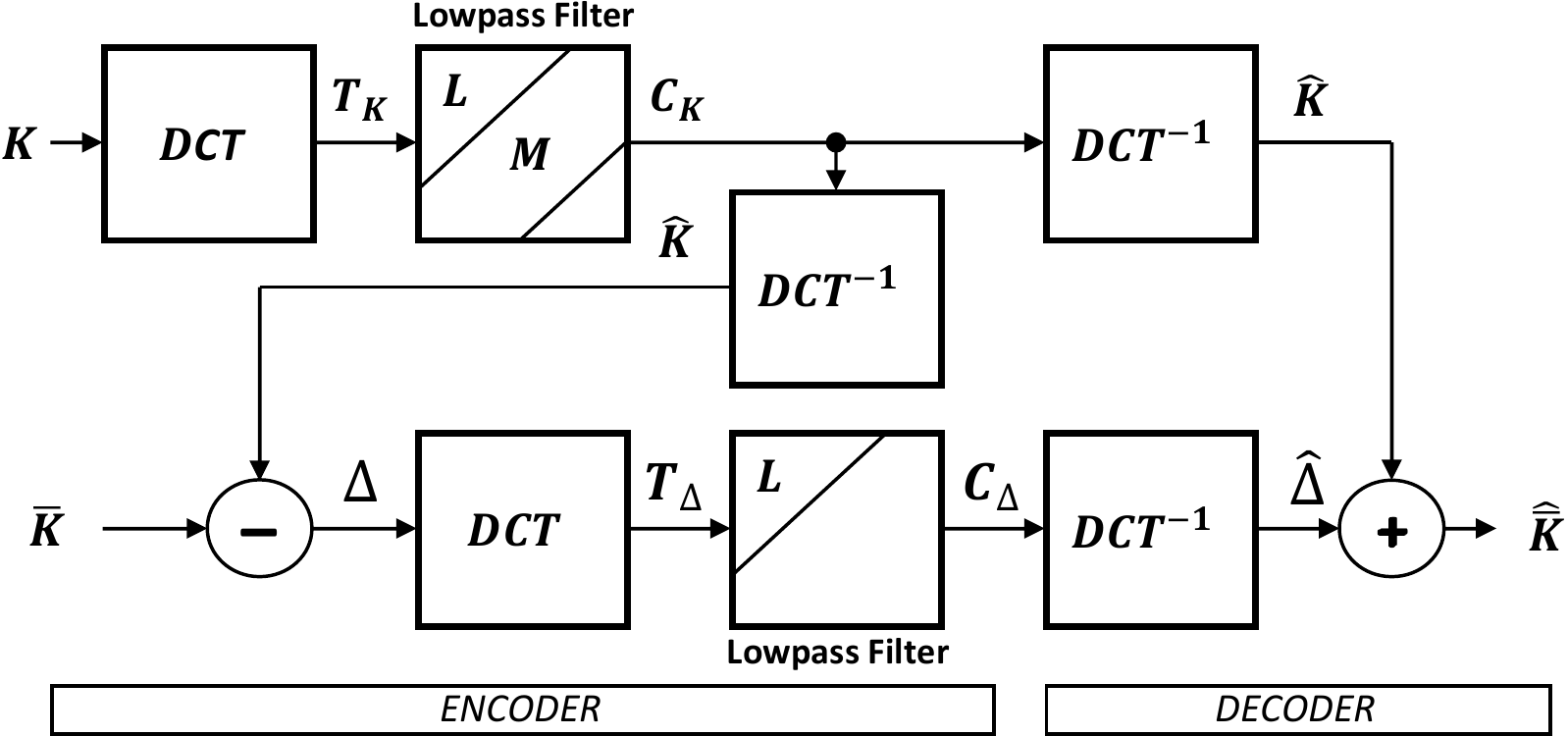}
    \caption{Non-distributed hybrid DPCM/DCT coder.}
    \label{DPCM}
\end{figure}

\subsection{Differential pulse code modulation} \label{dpcm}
Temporal DPCM exploits correlation between frames. Consider the hybrid DPCM/DCT codec in figure \ref{DPCM}. A $B \times B$ pixel block from key-frame $\mathbf{K}$ is transformed using a 2D DCT operation generating $B \times B$ transform coefficients $\mathbf{T_K}$. We define the compressive L-DCT-ZZ operation, as that operation that retains $(L+M)$ lowpass transform coefficients in JPEG \cite{jpeg} zigzag order. $L$ is the number of transform coefficients retained in a non-key frame and $M$ are the number of extra mid-band DCT coefficients retained in a key-frame, over and above those in the non-key frame. The resulting $\mathbf{C_K}$ transform coefficients are transmitted to the decoder, where an inverse 2D DCT reconstructs the key frame estimate $\mathbf{\hat{K}}$.

To exploit the correlation between a key frame block and a non-key frame block $\mathbf{\bar{K}}$, the key frame block estimate $\mathbf{\hat{K}}$ is subtracted from the non-key frame block. The resulting difference block $\Delta$ is transformed using the 2D DCT, generating transform coefficients $\mathbf{T}_\Delta$ which are lowpass filtered, retaining only $\mathbf{L}$ upper triangular transform coefficients $\mathbf{C}_\Delta$. At the decoder, $\mathbf{C}_\Delta$ is inverse transformed to reconstruct an estimate of the difference block $\hat{\Delta}$ that is added to the frame block estimate $\mathbf{\hat{K}}$ to estimate the non-key frame block $\mathbf{\hat{\bar{K}}}$.

Let the key frame pixel block $\mathbf{K}$ be arranged into a column vector $\mathbf{k}$. The linear unitary 2D DCT transform $\mathbf{D}$ then transforms $\mathbf{k}$ into the column vector of TCs $\mathbf{t_K}$
\begin{equation}
    \mathbf{t_K = Dk}
\end{equation}
where $\mathbf{D}$ is the $N \times N$ transform matrix and $N=B \times B$. The lowpass filtering operation is accomplished by element-wise multiplication (denoted by $\odot$) of $\mathbf{t_K}$ with a matrix consisting of $L+M$ 1's in the upper left hand triangle and 0's elsewhere, arranged as a vector $\mathbf{f_{L+M}}$, such that the L-DCT-ZZ vectorized coefficients of the key frame $\mathbf{c_K}$ are given by
\begin{equation} \label{ck}
\begin{split}
    \mathbf{c_K} & = \mathbf{t_K  \odot f_{L+M}} \\ 
                 & = \mathbf{(Dk) \odot f_{L+M}} \\
                 & = \mathbf{(Dk) \odot f_{L} + (Dk) \odot f_{M} }                 
    \end{split}
\end{equation}
The L-DCT-ZZ vectorized coefficients of the difference block are given by
\begin{equation} \label{cdelta}
\begin{split}
    \mathbf{c}_\Delta & = \mathbf{[D(\bar{k} - D^{-1}c_K)] \odot f_L} \\
                      & = \mathbf{[D\bar{k} - c_K] \odot f_L} \\
                      & = \mathbf{(D\bar{k}) \odot f_L - (Dk) \odot f_{L+M} \odot f_L} \\                    
                      & = \mathbf{(D\bar{k}) \odot f_L - (Dk) \odot f_L}
\end{split}
\end{equation}
since $ \mathbf{ f_{L+M} = f_L + f_M } $ and $\mathbf{ (f_L + f_M) \odot f_L = f_L }$, given that $\mathbf{f_L}$ and $\mathbf{f_M}$ are disjoint ($\mathbf{f_L \cap f_M = \{\overrightarrow{0}\}}$).

Equation (\ref{ck}) shows that the key frame entails transmitting the $L$ lowpass and $M$ mid-band L-DCT-ZZ coefficients. Equation (\ref{cdelta}) shows that the DPCM difference transform coefficients $\mathbf{c}_\Delta$ are composed of the lowpass coefficients of the non-key frame from which are subtracted from the lowpass coefficients of the key frame. Therefore, the encoder can be simplified to just two L-DCT-ZZ encoders one for the key frames retaining $(L+M)$ transform coefficients and one for the non-key frames retaining just $L$ coefficients. The difference block coefficients $\mathbf{c}_\Delta$ can then be calculated at the decoder by subtracting the key frame lowpass coefficients $\mathbf{(Dk) \odot f_L}$ from the non-key frame lowpass coefficients $\mathbf{(D \bar{k}) \odot f_L}$.

\subsection{Real-time reconstruction}
The VAL-DD encoder transmits the transform coefficients and the number $m_i$ of transform coefficients per block to the decoder. For the key frames, the decoder then reconstructs each image block using the inverse linear 2D-DCT. The reconstructed key block arranged as a column vector is given by
\begin{equation}
    \mathbf{\hat{k} = D^{-1}c_k}
\end{equation}
This can be achieved in real-time using matrix multiplication or fast implementations of the inverse 2D-DCT.

\begin{figure}[t]
    \centering
    \includegraphics[width=0.45\textwidth]{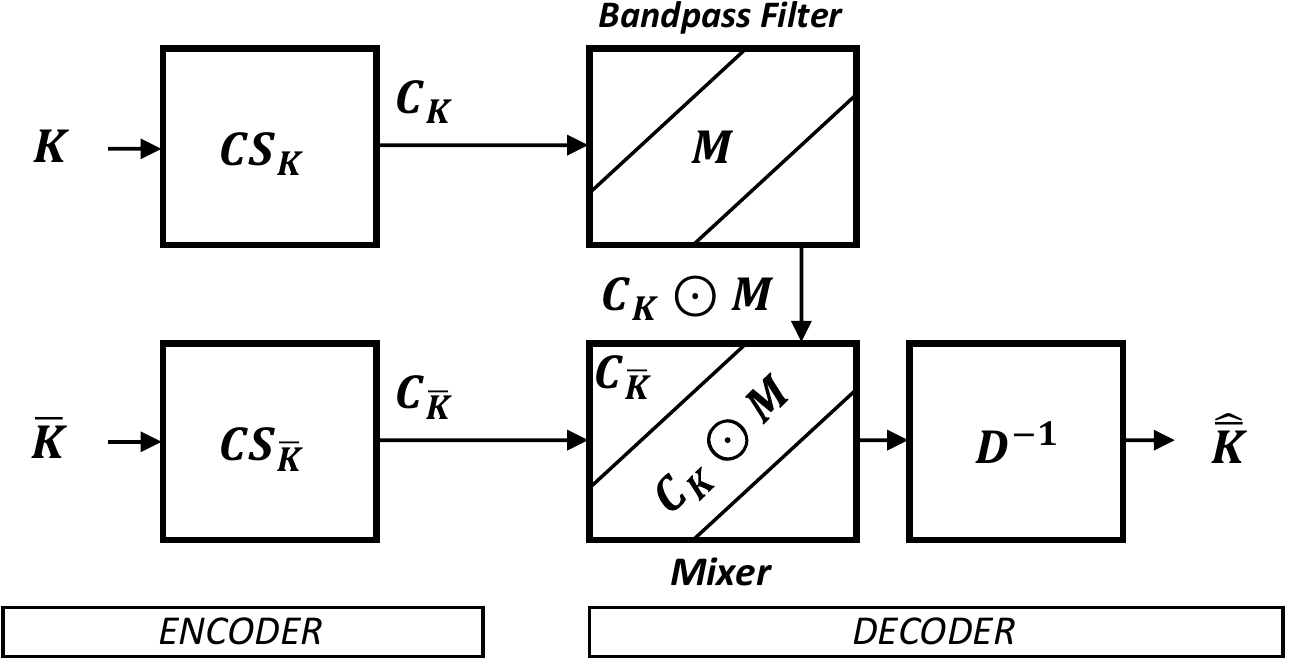}
    \caption{Distributed compressive DPCM in the DCT domain. $\mathbf{CS_K}$ and $\mathbf{CS_{\bar{K}}}$ represent the CS operations on the key and non-key blocks.}
    \label{DVC}
\end{figure}

The non-key frame can be computed by first calculating $\mathbf{c}_\Delta$ at the encoder as described in \ref{dpcm}, multiplying it by the inverse 2D DCT matrix $\mathbf{D^{-1}}$ to generate the difference block and then adding it to key frame. In vector form
\begin{equation}
    \mathbf{\hat{\bar{k}} = D^{-1}c}_\Delta + \mathbf{\hat{k}}
\end{equation}

However, we can also write
\begin{equation}
\begin{split}
    \mathbf{\hat{\bar{k}}} & = \mathbf{D^{-1}(c}_\Delta + \mathbf{c_K)} \\
                           & = \mathbf{D^{-1}[(D \bar{k}) \odot f_{L} + (Dk) \odot f_{M}]} 
\end{split}
\end{equation}
Thus the non-key frame can also be computed by first mixing (adding) the mid-band transform coefficients of the key frame to the lowpass transform coefficients of the non-key frame and calculating the inverse 2D DCT transform as shown in the distributed version of the DPCM operation in figure \ref{DVC}. Note that in figure \ref{VAL-VFI-fig}, $\mathbf{F_{n+1}}$ corresponds to key frames $\mathbf{K}$ in figure \ref{DVC}, and $\mathbf{F_{n+1,i}}$ to non-key frames $\bar{\mathbf{K}}$ for $i>0$. 

\subsection{Best pixel discriminator}
In unchanging parts of a frame, selecting key-frame blocks results in a higher-quality rendition. VFI results in estimating high-quality key-frame blocks and the subsequent DPCM process combines them with the low-pass current blocks. If the VFI is not accurate, image artefacts appear in the output blocks and this occurs where the image is changing most dynamically.

Following reconstruction of the current non-key frame and VFI from the key frame, we have three versions of the reconstructed frame; $\mathbf{\bar{K}_{n,i}}$ is low resolution non-key frame with no artefacts, $\mathbf{R}_{n,i}$ is the higher resolution VFI predicted frame but with the possibility of image artefacts, and $\mathbf{D_{n,i}}$ which is the DPCM output that may also have visible artefacts. To reduce the visibility of VFI artefacts, one option would be to average the two reconstructions together and this will indeed result in a better PSNR and MS-SSIM. 

A second option is to compare the pixels at location $(i,j)$ in both $\mathbf{R}_{n,i}$ and $\mathbf{D_{n,i}}$ and if the modulus of the difference $\lvert \mathbf{R_{n,i}}-\mathbf{D_{n,i}} \rvert$ is less than a threshold $T_D$ then the output pixel is set to $\mathbf{D_{n,i}}$; otherwise it is set to $\mathbf{\bar{K}_{n,i}}$.

This selection operation was found to improve the output quality considerably. The value of $T_D$ was investigated empirically. A value of $T_D=25$ was found to give good results with this best pixel discriminator (BPD) block in the VAL-VFI versions of our algorithms, whereas $T_D=10$ gave the best results with VAL-IDA-VFI.

The BPD results in lowpass pixels at the edges of moving objects in the current output frame. However the visual quality is still better as the human vision system is less sensitive to errors at moving boundaries.

\subsection{VAL-VFI parameters}

The VAL-VFI framework is characterised by a number of parameters that have to be taken into consideration, ideally optimized, namely: DCT block size; GOP size; compression ratio (or subrate) $\delta = M/N$, for key and non-key frames; IDA damping factor and iterations, if used; and BPD threshold $T_D$. 

\begin{table*}[ht]
\centering
\begin{normalsize}

\resizebox{\textwidth}{!}{%
\begin{tabular}{@{}lcccccccc@{}}
\toprule
\multicolumn{1}{c}{\multirow{2}{*}{\begin{tabular}[c]{@{}c@{}}Video\\ Sequence\end{tabular}}} &
  ME/MC &
  RRS &
  SGS-OF &
  VAL-VFI &
  VAL-IDA-VFI &
  VAL-IDA-VFI &
  VAL-VFI &
  VAL-IDA-VFI \\ \cmidrule(l){2-2} \cmidrule(l){3-3} \cmidrule(l){4-4} \cmidrule(l){5-5} \cmidrule(l){6-6} \cmidrule(l){7-7} \cmidrule(l){8-8} \cmidrule(l){9-9}  
\multicolumn{1}{c}{} &
  \begin{tabular}[c]{@{}c@{}}GOP=8 \\ $\delta_K = 0.7$\end{tabular} &
  \begin{tabular}[c]{@{}c@{}}GOP=8\\ $\delta_K = 0.7$\end{tabular} &
  \begin{tabular}[c]{@{}c@{}}GOP=8,\\ $\delta_K = 0.7$\end{tabular} &
  \begin{tabular}[c]{@{}c@{}}GOP=8\\ $\delta_K = 0.7$\end{tabular} &
  \begin{tabular}[c]{@{}c@{}}GOP=8\\ $\delta_K = 0.7$\end{tabular} &
  \begin{tabular}[c]{@{}c@{}}GOP=8\\ $\delta_K = 0.4$\end{tabular} &
  \begin{tabular}[c]{@{}c@{}}GOP=4\\ $\delta_K = 0.5$\end{tabular} &
  \begin{tabular}[c]{@{}c@{}}GOP=4\\ $\delta_K = 0.5$\end{tabular} \\ \midrule
Paris &
  29.63, 0.9872 &
  27.07, 0.9853 &
  32.00, 0.9946 &
  32.28, 0.9933 &
  \textbf{33.82}, 0.9932 &
  31.79, 0.9938 &
  31.26, 0.9941 &
  33.72, \textbf{0.9953} \\
Foreman &
  32.85, 0.9829 &
  36.51, 0.9943 &
  \textbf{37.20}, \textbf{0.9952} &
  35.45, 0.9903 &
  37.19, 0.9891 &
  37.10, 0.9924 &
  35.51, 0.9916 &
  36.61, 0.9920 \\
Coastguard &
  30.14, 0.9470 &
  31.20, 0.9535 &
  32.40, 0.9692 &
  32.35, 0.9680 &
  32.70, 0.9670 &
  32.10, 0.9745 &
  34.31, 0.9832 &
  \textbf{34.49}, \textbf{0.9833} \\
Hall &
  41.12, \textbf{0.9960} &
  35.68, 0.9947 &
  40.30, 0.9953 &
  42.02, 0.9959 &
  \textbf{42.21}, 0.9946 &
  41.19, 0.9951 &
  41.05, 0.9959 &
  41.78, 0.9959 \\
Mobile &
  21.39, 0.9059 &
  23.96, 0.9610 &
  27.13, 0.9897 &
  28.33, 0.9900 &
  28.75, 0.9896 &
  27.33, 0.9890 &
  27.82, 0.9903 &
  \textbf{29.45}, \textbf{0.9929} \\
News &
  38.39, 0.9967 &
  34.11, 0.9965 &
  39.88, \textbf{0.9978} &
  40.91, 0.9974 &
  \textbf{42.18}, 0.9973 &
  41.11, \textbf{0.9978} &
  40.77, 0.9977 &
  41.39, \textbf{0.9978} \\ \midrule 
\multicolumn{1}{c}{Average} &
  32.26, 0.9693 &
  31.42, 0.9809 &
  34.82, 0.9903 &
  35.22, 0.9892 &
  36.14, 0.9885 &
  35.10, 0.9904 &
  35.12, 0.9921 &
  \textbf{36.24}, \textbf{0.9929} \\ \bottomrule
\end{tabular}%
}
\end{normalsize}

\caption{VAL-VFI and VAL-IDA-VFI algorithms compared with ME/MC \cite{2011-Chen}, RRS \cite{2016-zhao} and SGS-OF \cite{2020-chen}. PSNR on the left, SSIM on the right. Best results in bold.}
\label{Table-VAL-VFI-1}

\end{table*}


\begin{figure*}
    \centering
    \includegraphics[width=0.9\textwidth]{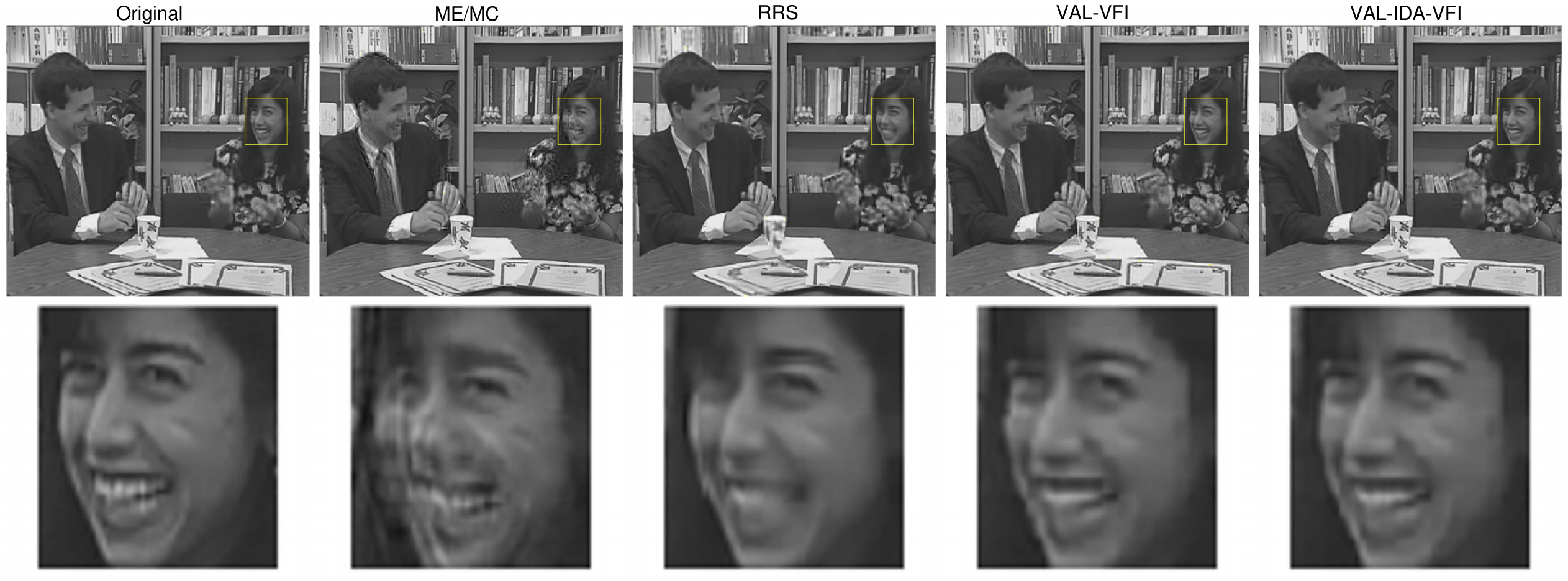}
    \caption{Frame 14 of the Paris sequence output Original, and reconstructed by ME/MC, RRS, VAL-VFI and VAL-IDA-VFI.}
    \label{Side-by-Side}
\end{figure*}

\begin{figure*}[t]
    \centering
    \includegraphics[width=0.9\textwidth]{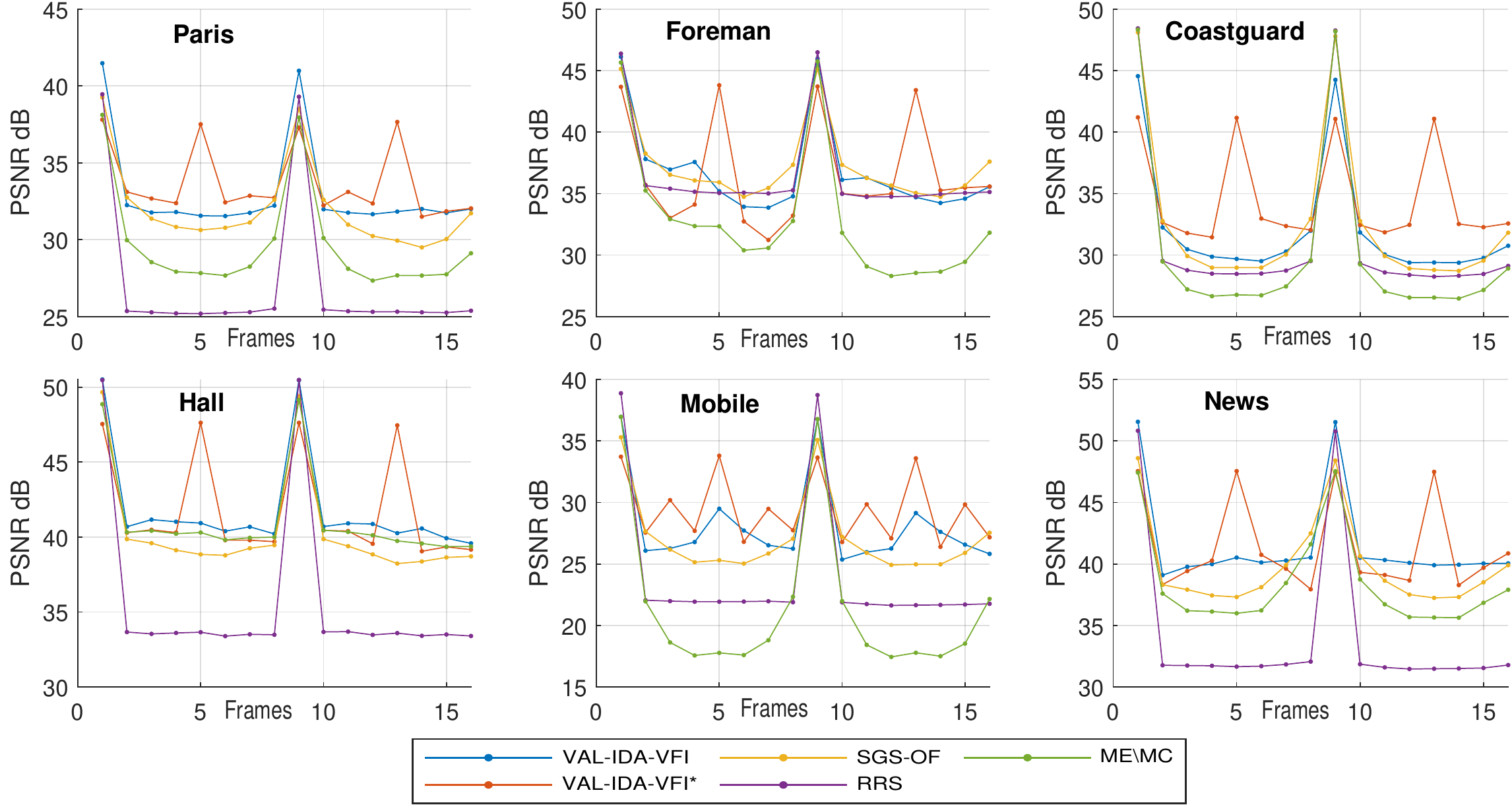}
    \caption{Comparing the PSNR for the first 16 frames of VidSet6 compressed with VAL-IDA-VFI, ME/MC, RRS and SGS-OF, with GOP = 8, VAL-IDA-VFI* with GOP=4. The average compression ratio is 0.175 across the GOP.}
    \label{VAL-Compare-PSNR}
\end{figure*}

\section{Simulation results} \label{VAL-Results}

The simulations in this section were executed on a server equipped with an Intel Xeon CPU E5-160 v3 clocked at 3.50 GHz, with 32GB of RAM, running MATLAB version 2019a on Linux 18.04. IDA requires the D-AMP-Toolbox from \cite{2017-DAMP}. The DnCNN code and models were downloaded from \cite{2017-DnCNN} and require the MatConvNet \cite{2015-vedaldi} package from \cite{2017-MatConvNet}. The DAIN code was downloaded from the author's site \cite{2020-DAIN}. We compiled it with Pytorch version 1.1 and it runs under Python 3.7.9. The VAL-IDA code exchanges Key and non-Key frames with DAIN at run time, using a file-based interface. The source code for this paper is available at: https://github.com/jzamm/val. Note that the algorithms presented in this paper extend to higher resolution images, the included code can be utilized without any modifications. The image test sets below were chosen to compare with published work, claiming SOTA performance, for which the source code was not available.

The VAL-VFI and VAL-IDA-VFI algorithms were compared with three algorithms in the literature; ME/MC \cite{2011-Chen}, RRS \cite{2016-zhao} and SGS-OF \cite{2020-chen}.  Six popular CIF video sequences were used for testing as in \cite{2020-chen}. We refer to this set as VidSet6. The first seventeen frames of VidSet6 were used with the seventeenth frame used by VFI to compute key frames in the second GOP, but PSNR and MS-SSIM \cite{2003-wang} results are only reported for the luminance (Y) plane of the first sixteen frames. 
The simulation uses blocks of size B = 16. The key compression ratio $\delta_K$, non-key compression ratio $\delta_{\bar{K}}$ and GOP size $G$ are related by the following equation:
\begin{equation} \label{Eqn-Deltas}
\frac{\delta_K + (G-1)\delta_{\bar{K}}}{G} = 0.175
\end{equation}

This ensures a constant average number of CS measurements per GOP. $G$, $\delta_K$ and $\delta_{\bar{K}}$ were initially set to 8, 0.7 and 0.1 respectively and then varied to optimize PSNR and MS-SSIM results as indicated below. Note that the higher the compression ratio, the more measurements are collected by the compressive sensing.

The original ME/MC code from \cite{2012-fowler-code} and RRS code from \cite{2018-Zhao} were modified to set the measurement matrix to the same DCT measurement matrix used by our algorithms. The use of the low-pass DCT measurement matrix improves the results of both ME/MC and RRS over the respective published results. SGS-OF code was not available and therefore, the SGS-OF results in table \ref{Table-VAL-VFI-1} are from the original paper in \cite{2020-chen}.

\subsection{PSNR and MS-SSIM}
The average PSNR and MS-SSIM results of VAL-VFI and VAL-IDA-VFI are compared with ME/MC, RRS and SGS-OF in table \ref{Table-VAL-VFI-1}. When the GOP size is 8, and the key and non-key frame compression ratios are 0.7 and 0.1 respectively, VAL-VFI exceeds the performance of ME/MC, RRS and SGS-OF by 2.96 dB, 3.80 dB and 0.40 dB respectively. It also exceeds the MS-SSIM of ME/MC and RRS by 0.0199 and 0.0083, but is 0.0011 short of SGS-OF. IDA reconstruction improves the VAL-VFI PSNR by 0.92 dB but decreases MS-SSIM marginally by 0.0007.

The GOP and $\delta_K$ values were then varied to optimise the VAL-VFI and VAL-IDA-VFI performance. When GOP=8, $\delta_K=0.4$ and $\delta_{\bar{K}}=0.1429$ as computed by equation (\ref{Eqn-Deltas}), VAL-IDA-VFI exceeds the state-of-the-art PSNR and MS-SSIM performance of SGS-OF by 0.28 dB and 0.0001. If the GOP size is reduced to 4, $\delta_K=0.5$ and $\delta_{\bar{K}}=0.0667$, the VAL-VFI PSNR and MS-SSIM exceed that of SGS-OF by 0.30 dB and 0.0018. VAL-IDA-VFI with GOP=4, $\delta_K=0.5$ and $\delta_{\bar{K}}=0.0667$ achieves our best results and improves VAL-VFI PSNR and MS-SSIM by 1.12 dB and 0.0008 respectively.

Figure \ref{VAL-Compare-PSNR} compares the PSNR of the Y component of the first 16 frames of the VidSet6 sequences. The GOP size is 8 in all cases except for VAL-IDA-VFI* for which the GOP size is 4. The compression ratio of the key frame is 0.7 for ME/MC, RRS and SGS-OF whereas it is 0.6 for VAL-IDA-VFI and 0.5 for VAL-IDA-VFI*. The average compression ratio over the whole GOP is 0.175 in all cases. As can seen from the figure, the non-key PSNR for VAL-IDA-VFI* is superior to the other algorithms in the Paris, Coastguard and Mobile sequences. VAL-IDA-VFI has the best non-key PSNR in the Hall and News sequences whereas SGS-OF prevails in the Foreman sequence. The variability of the PSNR is least for VAL-IDA-VFI* with the key compression ratio equal to 0.5.
\subsection{Visual quality and execution time} \label{VQ-and-Running-time}
The visual quality of the fourteenth frame of the Paris sequence produced by the VAL-VFI and VAL-IDA-VFI algorithms is compared with the original frame and the output of ME/MC and RRS algorithms in figure \ref{Side-by-Side}. VAL-VFI and VAL-IDA-VFI (GOP = 4, $\delta_K = 0.5, \delta_{\bar{K}} = 0.0667$) both have better rendition than ME/MC and RRS, with no artefacts like ME/MC and with sharper rendition. VAL-IDA-VFI improves on the VAL-VFI quality.
The non-optimized VAL-VFI reconstructs a frame in around 190ms, VAL-IDA-VFI in 2.1 seconds (20 iterations of the DnCNN algorithm, PSNR = 36.15 dB, MS-SSIM = 0.9928), ME/MC in 15s and RRS in 557s. SGS-OF is reported in \cite{2020-chen} to reconstruct a frame in 15s on a server with equivalent processing power.



\section{Conclusion}

We have proposed the VAL-VFI and VAL-IDA-VFI algorithms that exploit adaptive, block-based compressive sensing of video frames in the spatial domain, using deterministic DCT matrices. The reconstruction quality of the compressively sensed frames can be enhanced using an iterative denoising algorithm.

Our algorithms exploit temporal correlation at the encoder using the MDD adaptivity estimation algorithm to match the compressive sensing ratio with the underlying block sparsity. At the decoder we exploit temporal correlation using a GOP structure and a video frame interpolation CNN to predict non-key frames from higher-quality key frames. The quality of the VFI frames is then enhanced by performing temporal DPCM at the decoder. Finally, a best pixel discriminator can select the best pixel from the DPCM output or the reconstructed non-key frame, depending on the pixel error between the VFI prediction and the DPCM reconstruction. Simulation results show that our algorithms achieve state-of-the-art performance. The improvement in performance is shown in figure \ref{PSRN-vs-Time}.

Future work can study VAL-VFI in an end-to-end transmission system, digitally encoding the adaptive measurements prior to transmission.


\bibliographystyle{IEEEtran}
\bibliography{IEEEabrv, ref}
\begin{IEEEbiography}[{\includegraphics[width=1in,height=1.25in,clip,keepaspectratio]{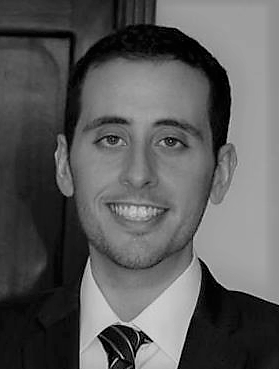}}]{Joseph  Zammit} received the B.Sc. degree in Communications and Computer Engineering  from the University of Malta, Msida, Malta, in 2012, the M.Sc. degree in Wireless and Optical Communications, from University College London, London, U.K., in 2013, and is a Ph.D. candidate at the Computer Laboratory, University of Cambridge, Cambridge, U.K. His research interests include sparse representations, wireless networks, deep learning, and distributed systems.
\end{IEEEbiography}
\begin{IEEEbiography}[{\includegraphics[width=1in,height=1.25in,clip,keepaspectratio]{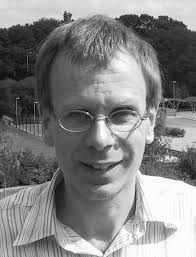}}]{Ian J. Wassell} received the B.Sc. and B.Eng. degrees from the University of Loughborough, Loughborough, U.K., in 1983, and the Ph.D. degree from
the University of Southampton, Southampton, U.K., in 1990. He is a Senior Lecturer in the Computer Laboratory, University of Cambridge, Cambridge, U.K. He has in excess of 15 years experience in the simulation and design of radio communication systems gained via a number of positions in industry and higher education. He has published more than 180 papers concerning wireless communication systems. His research interests include fixed wireless access, sensor networks, cooperative networks, propagation modeling, compressive sensing, and cognitive radio. He is a member of the IET and a Chartered Engineer.
\end{IEEEbiography}
\vfill
\EOD
\end{document}